\documentclass[onecolumn]{ceurart}

\selectcolormodel{natural}
\usepackage{ninecolors}
\selectcolormodel{rgb}
\usepackage{tabularray}

\begin{document}

\copyrightyear{2024}
\copyrightclause{Copyright for this paper by its authors.
  Use permitted under Creative Commons License Attribution 4.0
  International (CC BY 4.0).}

\conference{Proceedings of the Best BPM Dissertation Award, Doctoral Consortium, and Demonstrations \& Resources Forum co-located with 22nd International Conference on Business Process Management (BPM 2024), Krakow, Poland, September 1st to 6th, 2024.}

\title{BPMN Analyzer 2.0: Instantaneous, Comprehensible, and Fixable Control Flow Analysis for Realistic BPMN Models}

\author[1]{Tim Kräuter}
[email=tkra@hvl.no]
\author[1]{Patrick Stünkel}
[email=past@hvl.no]
\author[1]{Adrian Rutle}
[email=aru@hvl.no]
\author[1]{Yngve Lamo}
[email=yla@hvl.no]
\author[2,1]{Harald König}
[email=harald.koenig@fhdw.de]
\address[1]{Western Norway University of Applied Sciences, Bergen, Norway}
\address[2]{FHDW Hannover, Germany}

\begin{abstract}
Many business process models contain control flow errors, such as deadlocks or livelocks, which hinder proper execution.
In this paper, we introduce a new tool that can instantaneously identify control flow errors in BPMN models, make them understandable for modelers, and suggest corrections to resolve them.
We demonstrate that detection is instantaneous by benchmarking our tool against synthetic BPMN models with increasing size and state space complexity, as well as realistic models.
Moreover, the tool directly displays detected errors in the model, including an interactive visualization, and suggests fixes to resolve them.
The tool is open source, extensible, and integrated into a popular BPMN modeling tool.
\end{abstract}


\maketitle

\section{Introduction}

Business Process Modeling Notation (BPMN) is becoming increasingly popular for automating processes and orchestrating people and systems.
However, many process models suffer from control flow errors, such as deadlocks, livelocks, and starvation~\cite{fahlandAnalysisDemandInstantaneous2011}.
These errors hinder the correct execution of BPMN models and may be detected late in the development process, resulting in elevated costs.

\begin{figure}[ht]
	\centering
	\includegraphics[width=0.6\linewidth]{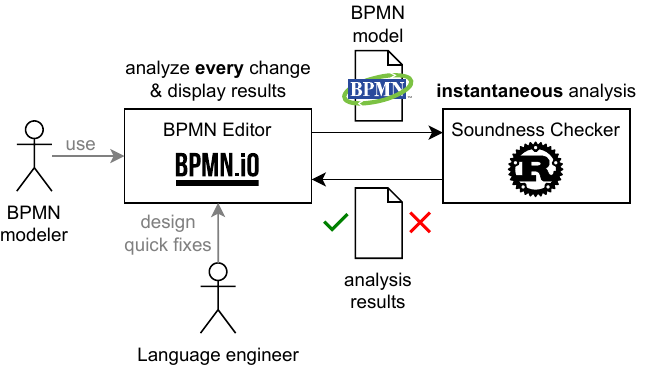}
	\caption{Overview of the BPMN Analyzer 2.0}
	\label{fig:overview}
\end{figure}

In this paper, we describe a new tool, the \textit{BPMN Analyzer 2.0}\footnote{
In the following, we will use BPMN Analyzer to refer to the BPMN Analyzer 2.0.}, for analyzing BPMN process models to detect control flow errors \textit{already} during modeling.
\autoref{fig:overview} shows an overview of the tool.
The UI is based on the popular \textit{bpmn.io} ecosystem, while the analysis is implemented in \textit{Rust} for optimal performance and memory efficiency.
We perform a breadth-first state space exploration to check soundness and safeness~\cite{corradiniClassificationBPMNCollaborations2018} \textit{on the fly} to uncover control flow errors.
Consequently, the tool can detect deadlocks, livelocks, starvation, dead activities, and lack of synchronization in BPMN models.
The BPMN Analyzer is open source and accessible online alongside a video demonstration\footnote{Tool: \url{https://timkraeuter.com/bpmn-analyzer-js/}, Video: \url{https://www.youtube.com/watch?v=Nv2W-hXNZYA}}~\cite{krauterInstantaneousComprehensibleFixable2024}.

The tool can check models after each change since analysis is \textit{instantaneous} according to~\cite{fahlandAnalysisDemandInstantaneous2011}, i.e., it takes 500ms or less.
Furthermore, we ensure the results are \textit{comprehensible} by highlighting possible violations directly in the model and displaying an interactive counterexample visualization.
Finally, the tool suggests \textit{fixes} for the most common control flow errors and can be extended to suggest more fixes in the future.

Fahland et al.~\cite{fahlandAnalysisDemandInstantaneous2011} describe \textit{coverage}, \textit{immediacy}, and \textit{consumability} as the main challenges for users unaccustomed to formal analysis.
The BPMN Analyzer addresses all these challenges since it supports the most common BPMN elements used in practice (coverage), provides \textit{instantaneous} results (immediacy), and a \textit{comprehensible} user interface (consumability), even including suggestions for fixes.
Developers of industrial BPMN software also like our tool, especially the End-to-end user journey~\cite{krauterInstantaneousComprehensibleFixable2024}.
Thus, this supports our claim that the UI is understandable for users unfamiliar with formal analysis.

In the remainder of the paper, we describe how instantaneous, comprehensible, and fixable control flow error detection is achieved in \autoref{sec:innovations}.
Then, we discuss tool maturity in \autoref{sec:maturity} before concluding in \autoref{sec:conclusion}.

\section{Innovations} \label{sec:innovations} 
The BPMN Analyzer has three main innovations: \textbf{instantaneous}, \textbf{comprehensible}, and \textbf{fixable} control flow error detection.
In this section, we will present the innovations, and more details can be found in our extended paper~\cite{krauterInstantaneousComprehensibleFixable2024}.

\subsection{Instantaneous Analysis}

We demonstrate instantaneous control flow analysis by benchmarking our tool in \textit{three} scenarios.
For all our benchmarks, we use the hyperfine benchmarking tool (version 1.18.0), which calculates the average runtime when executing each control flow analysis ten or more times.
We ran the benchmarks on Ubuntu 22.04.4 with an AMD Ryzen 7700X processor (4.5GHz) and 32 GB of RAM (5600 MHz).
All used BPMN models, our tools to generate them, and benchmarking scripts to run them are available in~\cite{krauterInstantaneousComprehensibleFixable2024}.

\textbf{First}, we benchmarked how our tool handles \textbf{BPMN models of growing size}.
We generated 500 synthetic BPMN models starting with five elements up to 4000.
The models repeatedly contain three activities and an exclusive/parallel block with two branches containing one activity per branch (see ~\cite{krauterInstantaneousComprehensibleFixable2024}).
The BPMN Analyzer spends from 1 ms up to 9 ms for the BPMN models~\cite{krauterInstantaneousComprehensibleFixable2024} compared to 0.7 s up to 14 s in our previous tool~\cite{krauterHigherorderTransformationApproach2024}.
In summary, the runtime grows linearly with the state space.

\textbf{Second}, we benchmarked the tool against a synthetic data set of models that led to a state space explosion.
This represents a \textit{worst-case} scenario for formal analysis.
We generated a data set of models~\cite{krauterInstantaneousComprehensibleFixable2024} with a growing number of parallel branches with increasing length, like~\cite{corradiniFormalApproachAnalysis2021}.

\autoref{tab:parallel-branches-benchmark} shows the average runtime of our tool when analyzing these models.
The BPMN Analyzer explores the entire state space while simultaneously analyzing the control flow, i.e., verifying soundness properties.
The models' state space grows exponentially, leading to the same order of growth in runtime.
Our analysis is not instantaneous anymore when approaching 17 parallel branches of length 1 (see \autoref{tab:parallel-branches-benchmark}).
However, analysis is still instantaneous for more reasonable models with five parallel branches of length 5 or 3 branches of length 20.
Other tools report 2-3s of runtime for most soundness properties and 30s for a model with five parallel branches~\cite{corradiniFormalApproachAnalysis2021}, which took milliseconds in our tool.

\begin{table}[ht]
\scalebox{1}{ 
    \begin{minipage}{.5\linewidth}
	   \centering
	   \caption{Benchmark results of the parallel branches models}
	   \label{tab:parallel-branches-benchmark}
	   \SetTblrInner{colsep=2pt}
	   \begin{tblr}{
	   		column{1-X} = {c},
	   		column{Y-Z} = {r},
	   		hline{1, 2, Z} = {-}{1.2pt, solid}, 
	   		hline{8, 9} = {-}{dashed},
	   		vline{2-Y} = {2-Z}{solid}, 
	   	}
	       \textbf{Branches} & \textbf{Branch Length} &\textbf{Runtime} & \textbf{States} \\
	       5 & 1 & 1 ms& 35 \\
	       10 & 1 & 3 ms& 1.027 \\
	       15 & 1 & 161 ms& 32.771 \\
	       16 & 1 & 360 ms& 65.539 \\
	       17 & 1 & 790 ms& 131.075 \\
	       20 & 1 & 8.803 ms& 1.048.579 \\
	       5 & 5 & 14 ms& 7.779 \\
	       3 & 20 & 11 ms& 9.264 \\
	   \end{tblr}
    \end{minipage}
    \begin{minipage}{.5\linewidth}
	   \centering
	   \caption{Benchmark results of the realistic BPMN models}
	   \label{tab:realistic-models-benchmark}
	   \SetTblrInner{colsep=2pt}
	   \begin{tblr}{
	   		column{Y-Z} = {r},
	   		hline{1, 2, Z} = {-}{1.2pt, solid}, 
	   		hline{5} = {-}{dashed},
	   		vline{2-Y} = {2-Z}{solid}, 
	   	}
	   	   \textbf{Model name} &\textbf{Runtime} & \textbf{States} \\
	   	   e001~\cite{houhouFirstOrderLogicVerification2022} & 1 ms & 39 \\
	   	   e002~\cite{houhouFirstOrderLogicVerification2022} & 1 ms & 39 \\
	   	   e020~\cite{houhouFirstOrderLogicVerification2022} & 10 ms & 5356 \\
	   	   credit-scoring-async\footref{footnote:camundaResearch} & 1 ms & 60 \\
	   	   credit-scoring-sync\footref{footnote:camundaResearch} & 1 ms & 140 \\
	   	   dispatch-of-goods\footref{footnote:camundaResearch} & 1 ms & 103\\
	   	   recourse\footref{footnote:camundaResearch} & 1 ms & 77 \\
	   	   self-service-restaurant\footref{footnote:camundaResearch} & 1 ms & 190 \\
	   \end{tblr}
    \end{minipage} 
}
\end{table}

\textbf{Third}, we applied our tool to eight \textbf{realistic models}, where three models (e001, e002, e020) are taken from~\cite{houhouFirstOrderLogicVerification2022}, and the remaining five models are part of the Camunda BPMN for research repository\footnote{\url{https://github.com/camunda/bpmn-for-research} \label{footnote:camundaResearch}}.
\autoref{tab:realistic-models-benchmark} shows each model's average runtime and number of states.
The BPMN Analyzer takes 1-10ms for e001, e002, and e020~\cite{krauterInstantaneousComprehensibleFixable2024}, while~\cite{houhouFirstOrderLogicVerification2022} and~\cite{krauterHigherorderTransformationApproach2024} report 3.66-10.26s and 1-1.75s respectively.
The benchmarks in~\cite{krauterHigherorderTransformationApproach2024} were run on the same hardware, while the machine used in~\cite{houhouFirstOrderLogicVerification2022} was less powerful.
Our analysis is instantaneous for nearly all BPMN models since most have less than 1000 states, according to~\cite{fahlandAnalysisDemandInstantaneous2011}.

\subsection{Comprehensible Analysis}
We implemented two features to make control flow analysis understandable for everyone. 
\textbf{First}, we highlight the problematic elements that cause control flow errors by directly attaching red overlays to them in the model.
In addition, there is a summary panel in the top-right stating if any errors are found.

\textbf{Second}, we use \textit{tokens} to visualize errors \textit{interactively}, i.e., show an execution leading to the error.
Our analysis provides sample executions resulting in the found control-flow errors, which we visualize in the editor by showing how tokens move from the process start to an erroneous state.
We are unaware of other tools that visualize errors directly and allow interactions, such as stopping/resuming and restarting.

In \autoref{fig:screenshots}, the visualization has been \textit{paused} just before an \textit{unsafe} state was reached.
One token is already located at the marked sequence flow, while a second token is currently waiting at the exclusive gateway \textsf{e1}.
The visualization can be resumed or restarted using the play and restart button on the left side.
The gateway \textsf{e1} will execute when resumed, resulting in two tokens at the subsequent sequence flow, i.e., an unsafe execution state.
Furthermore, one can control the visualization speed using the bottom buttons next to the speedometer.

\subsection{Fixable Analysis}
Besides detecting, highlighting, and visualizing control flow errors, the BPMN Analyzer suggests fixes like \textit{quick fixes} in IDEs.
Quick fixes cannot be provided for all errors, but we currently cover many patterns leading to deadlocks, lack of synchronization, message starvation, and reused end events.
The quick fixes we support are described in detail in~\cite{krauterInstantaneousComprehensibleFixable2024} and can be extended independently of the formal analysis.
We are unaware of other tools that can fix identified control-flow errors.

\begin{figure}[ht]
	\centering
	\includegraphics[width=1\linewidth]{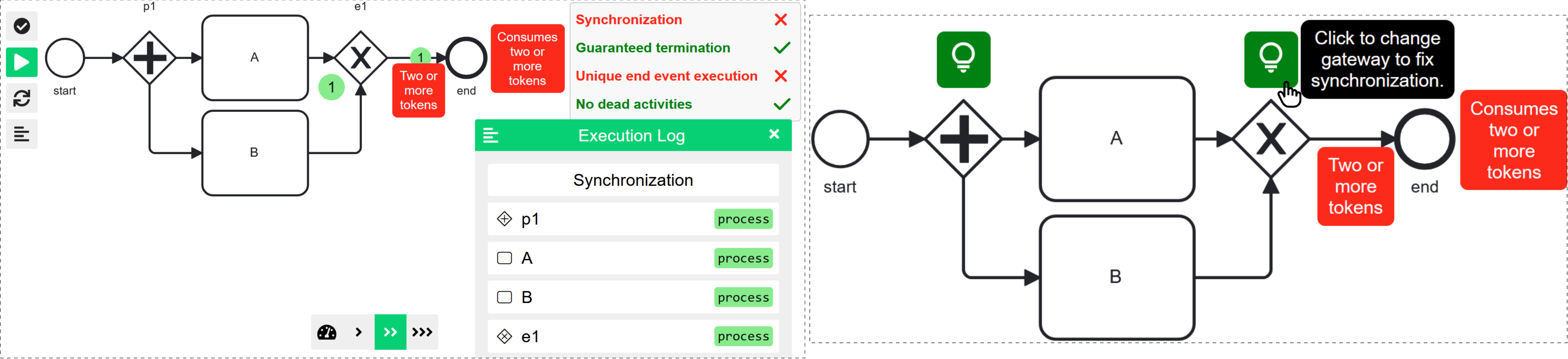}
	\caption{Execution visualization (left) and suggested \textit{quick fixes} (right) in the BPMN Analyzer}
	\label{fig:screenshots}
\end{figure}

For example, \autoref{fig:screenshots} shows a screenshot of our tool, where quick fixes are depicted as green overlays containing a light bulb icon.
A user can apply a quick fix by clicking on a green overlay and instantly see the changes regarding control flow errors.
If unhappy with the result, a user can undo all changes since each quick fix is entirely revertible due to the command pattern.
A user might not like a quick fix if it not only fixes an error but also has unintended side effects, such as introducing a different control flow error.

\section{Maturity of the Tool} \label{sec:maturity}
The BPMN Analyzer incorporates many findings from our previous work~\cite{krauterHigherorderTransformationApproach2024} while focusing on instantaneous and understandable error detection, as described in the previous section.
The tool is open source~\cite{krauterInstantaneousComprehensibleFixable2024}, and we ensure high code quality by employing industry best practices such as rigorous static analysis and testing.
Furthermore, we received positive feedback from companies in the BPMN process orchestration space~\cite{krauterInstantaneousComprehensibleFixable2024}.

\section{Conclusion \& Future Work} \label{sec:conclusion}

In this paper, we describe the novel \textit{BPMN Analyzer} that provides instantaneous, comprehensible, and fixable BPMN control flow error detection and is integrated into a popular BPMN modeling tool.
We benchmarked our tool against synthetic and realistic BPMN models to demonstrate instantaneous soundness checking.
We address the three main challenges, \textit{coverage}, \textit{immediacy}, and \textit{consumability}, to provide formal analysis to non-expert users as identified in~\cite{fahlandAnalysisDemandInstantaneous2011}.
In addition, our tool offers quick fixes for common patterns that lead to control flow errors.
One can understand the BPMN Analyzer as a BPMN-specific model checker, implemented in Rust paired with an intuitive user interface based on the popular \textit{bpmn.io} ecosystem that is open for extension by design.

In future work, we aim to improve our tool by providing more quick fixes, considering advanced BPMN elements such as different events, and ranking quick fixes based on usefulness and previous user behavior.
Finally, we aspire to test our tool in a real-world scenario to gather feedback and measure its impact on productivity.

\bibliography{bib}

\begin{thebibliography}{6}
\expandafter\ifx\csname natexlab\endcsname\relax\def\natexlab#1{#1}\fi
\providecommand{\url}[1]{\texttt{#1}}
\providecommand{\href}[2]{#2}
\providecommand{\path}[1]{#1}
\providecommand{\DOIprefix}{doi:}
\providecommand{\ArXivprefix}{arXiv:}
\providecommand{\URLprefix}{URL: }
\providecommand{\Pubmedprefix}{pmid:}
\providecommand{\doi}[1]{\href{http://dx.doi.org/#1}{\path{#1}}}
\providecommand{\Pubmed}[1]{\href{pmid:#1}{\path{#1}}}
\providecommand{\bibinfo}[2]{#2}
\ifx\xfnm\relax \def\xfnm[#1]{\unskip,\space#1}\fi
\bibitem[{Fahland et~al.(2011)Fahland, Favre, Koehler, Lohmann, V{\"o}lzer, and Wolf}]{fahlandAnalysisDemandInstantaneous2011}
\bibinfo{author}{D.~Fahland}, \bibinfo{author}{C.~Favre}, \bibinfo{author}{J.~Koehler}, \bibinfo{author}{N.~Lohmann}, \bibinfo{author}{H.~V{\"o}lzer}, \bibinfo{author}{K.~Wolf},
\newblock \bibinfo{title}{Analysis on demand: {{Instantaneous}} soundness checking of industrial business process models},
\newblock \bibinfo{journal}{Data \& Knowledge Engineering} \bibinfo{volume}{70} (\bibinfo{year}{2011}) \bibinfo{pages}{448--466}. \DOIprefix\doi{10.1016/j.datak.2011.01.004}.
\bibitem[{Corradini et~al.(2018)Corradini, Muzi, Re, and Tiezzi}]{corradiniClassificationBPMNCollaborations2018}
\bibinfo{author}{F.~Corradini}, \bibinfo{author}{C.~Muzi}, \bibinfo{author}{B.~Re}, \bibinfo{author}{F.~Tiezzi},
\newblock \bibinfo{title}{A {{Classification}} of {{BPMN Collaborations}} based on {{Safeness}} and {{Soundness Notions}}},
\newblock \bibinfo{journal}{Electronic Proceedings in Theoretical Computer Science} \bibinfo{volume}{276} (\bibinfo{year}{2018}) \bibinfo{pages}{37--52}. \DOIprefix\doi{10.4204/EPTCS.276.5}.
\bibitem[{Kr{\"a}uter et~al.(2024{\natexlab{a}})Kr{\"a}uter, St{\"u}nkel, Rutle, K{\"o}nig, and Lamo}]{krauterInstantaneousComprehensibleFixable2024}
\bibinfo{author}{T.~Kr{\"a}uter}, \bibinfo{author}{P.~St{\"u}nkel}, \bibinfo{author}{A.~Rutle}, \bibinfo{author}{H.~K{\"o}nig}, \bibinfo{author}{Y.~Lamo}, \bibinfo{title}{Instantaneous, {{Comprehensible}}, and {{Fixable Soundness Checking}} of {{Realistic BPMN Models}}}, \bibinfo{year}{2024}{\natexlab{a}}. \href{http://arxiv.org/abs/2407.03965}{{\tt arXiv:2407.03965}}.
\bibitem[{Kr{\"a}uter et~al.(2024{\natexlab{b}})Kr{\"a}uter, Rutle, K{\"o}nig, and Lamo}]{krauterHigherorderTransformationApproach2024}
\bibinfo{author}{T.~Kr{\"a}uter}, \bibinfo{author}{A.~Rutle}, \bibinfo{author}{H.~K{\"o}nig}, \bibinfo{author}{Y.~Lamo}, \bibinfo{title}{A higher-order transformation approach to the formalization and analysis of {{BPMN}} using graph transformation systems}, \bibinfo{year}{2024}{\natexlab{b}}. \href{http://arxiv.org/abs/2311.05243}{{\tt arXiv:2311.05243}}.
\bibitem[{Corradini et~al.(2021)Corradini, Fornari, Polini, Re, Tiezzi, and Vandin}]{corradiniFormalApproachAnalysis2021}
\bibinfo{author}{F.~Corradini}, \bibinfo{author}{F.~Fornari}, \bibinfo{author}{A.~Polini}, \bibinfo{author}{B.~Re}, \bibinfo{author}{F.~Tiezzi}, \bibinfo{author}{A.~Vandin},
\newblock \bibinfo{title}{A formal approach for the analysis of {{BPMN}} collaboration models},
\newblock \bibinfo{journal}{Journal of Systems and Software} \bibinfo{volume}{180} (\bibinfo{year}{2021}) \bibinfo{pages}{111007}. \DOIprefix\doi{10.1016/j.jss.2021.111007}.
\bibitem[{Houhou et~al.(2022)Houhou, Baarir, Poizat, Qu{\'e}innec, and Kahloul}]{houhouFirstOrderLogicVerification2022}
\bibinfo{author}{S.~Houhou}, \bibinfo{author}{S.~Baarir}, \bibinfo{author}{P.~Poizat}, \bibinfo{author}{P.~Qu{\'e}innec}, \bibinfo{author}{L.~Kahloul},
\newblock \bibinfo{title}{A {{First-Order Logic}} verification framework for communication-parametric and time-aware {{BPMN}} collaborations},
\newblock \bibinfo{journal}{Information Systems} \bibinfo{volume}{104} (\bibinfo{year}{2022}) \bibinfo{pages}{101765}. \DOIprefix\doi{10.1016/j.is.2021.101765}.

\end{thebibliography}

\end{document}